\documentclass[aps,prc,twocolumn,showpacs,preprintnumbers,amsmath,amssymb,superscriptaddress]{revtex4-1}

\usepackage{graphicx}% Include figure files
\usepackage{dcolumn}% Align table columns on decimal point
\usepackage{bm}% bold math
\usepackage{epsfig}% Include figure files

\begin{document}

\preprint{APS/123-QED}

\title{Astrophysical analysis of the measurement of ($\alpha$,$\gamma$) and ($\alpha$,n) cross sections of $^{169}$Tm}

\author{T. Rauscher}%
\email{Thomas.Rauscher@unibas.ch}
\affiliation{%
Department of Physics, University of Basel, CH-4056 Basel, Switzerland}%
\author{G. G. Kiss}%
\affiliation{%
Institute of Nuclear Research (ATOMKI), H-4001 Debrecen, POB.51., Hungary}%
\author{T. Sz\"ucs}%
\affiliation{%
Institute of Nuclear Research (ATOMKI), H-4001 Debrecen, POB.51., Hungary}%
\author{Zs.\ F\"ul\"op}%
\affiliation{%
Institute of Nuclear Research (ATOMKI), H-4001 Debrecen, POB.51., Hungary}%
\author{C. Fr\"ohlich}%
\affiliation{%
Department of Physics, North Carolina State University, Raleigh, NC 27695, USA}%
\author{Gy. Gy\"urky}%
\affiliation{%
Institute of Nuclear Research (ATOMKI), H-4001 Debrecen, POB.51., Hungary}%
\author{Z. Hal\'asz}%
\affiliation{%
Institute of Nuclear Research (ATOMKI), H-4001 Debrecen, POB.51., Hungary}%
\author{Zs. Kert\'esz}%
\affiliation{%
Institute of Nuclear Research (ATOMKI), H-4001 Debrecen, POB.51., Hungary}%
\author{E. Somorjai}%
\affiliation{%
Institute of Nuclear Research (ATOMKI), H-4001 Debrecen, POB.51., Hungary}%

\date{\today}

\begin{abstract}
Reaction cross sections of $^{169}$Tm($\alpha$,$\gamma$)$^{173}$Lu and $^{169}$Tm($\alpha$,n)$^{172}$Lu have been measured in the energy range $12.6\leq E_\alpha \leq 17.5$ MeV and $11.5\leq E_\alpha \leq 17.5$ MeV, respectively, using the recently introduced method of combining activation with X-ray counting. Improved shielding allowed to measure the ($\alpha$,$\gamma$) to lower energy than previously possible. The combination of ($\alpha$,$\gamma$) and ($\alpha$,n) data made it possible to study the energy dependence of the $\alpha$ width. While absolute value and energy dependence are perfectly reproduced by theory at the energies above 14 MeV, the observed change in energy dependence at energies below 14 MeV requires a modification of the predicted $\alpha$ width. Using an effective, energy-dependent, local optical $\alpha$+nucleus potential it is possible to reproduce the data but the astrophysical rate is still not well constrained at $\gamma$-process temperatures. The additional uncertainty stemming from a possible modification of the compound formation cross section is discussed. Including the remaining uncertainties, the recommended range of astrophysical reaction rate values at 2 GK is higher than the previously used values by factors of $2-37$.
\end{abstract}


\maketitle

\section{Astrophysical motivation}
\label{sec:mot}

Two neutron capture processes, the $s$ and $r$ process, are required to produce the bulk of natural nuclides above Fe \cite{b2fh,kapgall,arngorr}. These two processes cannot, however, create 35 neutron-deficient, stable, rare isotopes between Se and Hg, which are termed $p$ nuclei.
Photodisintegration of stable nuclei in the O/Ne shell of massive stars during a core-collapse supernova explosion has been suggested as a production mechanism for these nuclei \cite{arn,woohow,rayet95}. Such a so-called $\gamma$ process commences by sequences of ($\gamma$,n) reactions which are replaced by ($\gamma$,p) and ($\gamma$,$\alpha$) reactions when reaching sufficiently neutron-deficient nuclides in an isotopic chain \cite{raunic}.

Two mass regions have remained problematic when explaining the production of $p$ nuclei by the $\gamma$ process in core-collapse supernovae: the lightest $p$ nuclei with mass numbers $A<100$ and those in an intermediate region at $150\leq A \leq165$ are underproduced \cite{arngorp,woohow,rayet95,rhhw02,hegXX}. While the explanation of the light $p$ nuclei most likely requires a different astrophysical model, the problem in the intermediate mass region may still be solved by improved reaction rates.

For the $\gamma$ process, photodisintegrations happen in the plasma temperature range $2.0\leq T\leq 3.0$ GK. The temperature is tightly constrained by the necessity to photodisintegrate the lighter, more tightly bound seed nuclei while also retaining heavy $p$ nuclides. Several layers with slightly different temperatures contribute to $p$ nucleosynthesis in a star. The rates for heavier nuclei have to be known in the lower part of the temperature region because they would be destroyed completely at slightly higher temperatures. Moreover, ($\gamma$,$\alpha$) reactions have been found important for the intermediate and heavy mass region, whereas ($\gamma$,p) dominates in the lighter mass region of the $\gamma$-process \cite{raudeflect,rappgamma}. The photodisintegration reaction rates are usually computed from capture rates by applying the reciprocity principle of stellar rates \cite{fow74,raureview}. Therefore a measurement of $\alpha$ capture may determine also the photodisintegration rates, provided the g.s.\ contribution to the stellar rate is large. Due to the Coulomb barrier, the ($\alpha$,$\gamma$) reaction cross sections are tiny at astrophysical energy and thus currently unmeasurable. Going to as low energy as possible, however, already some discrepancies between data and predictions have been found in previous measurements. Low-energy $\alpha$ capture on heavy nuclei seems to be often overpredicted, although there is not yet enough data to draw a conclusive picture. It underlines, nevertheless, that the underproduction of p-nuclides in the range $150\leq A \leq165$ observed in stellar models may have a nuclear physics cause.

Investigations of rates for the $\gamma$ process are not only important for nucleosynthesis in core-collapse supernovae. Simulations of the thermonuclear explosion of a White Dwarf (type Ia supernova) also found $p$ nuclei being produced in a $\gamma$ process \cite{travaglio}. Regardless of the site, $\gamma$ process studies require a sound determination of the relevant astrophysical reaction rates by nuclear physics investigations. The $\gamma$-process reaction networks include hundreds of nuclei and thousands of reactions, mostly on unstable nuclei. All of the reaction rates are predicted in the Hauser-Feshbach statistical model of nuclear reactions \cite{haufesh,adndt}. Since ($\gamma$,p) and ($\gamma$,$\alpha$) reactions occur at unstable isotopes, current experimental techniques have to aim for testing reaction model predictions at stability and to provide the data for a global improvement of these models and their input.

The nucleus $^{169}$Tm is not a $p$ nuclide but it is close to the problematic mass range. Very few $\alpha$-induced reaction data are known in this mass range and none close to the astrophysically relevant energy range. This made $^{169}$Tm an interesting target for investigation using the newly introduced method of activation with subsequent X-ray counting, supplementing the convential $\gamma$-counting methods. Details of the experimental method and first results were already published in \cite{kis_plb,kis_npa}. Here we introduce the additional data, extending the ($\alpha$,$\gamma$) cross sections to lower energies, and focus on a discussion of the implications for constraining the astrophysical reaction rate for ($\alpha$,$\gamma$) at $\gamma$-process temperatures.

\section{Experimental procedure and results}
\label{sec:pro}

\begingroup
\squeezetable
\begin{table}
\caption{\label{tab:results_an} Extended experimental $S$ factors (and recently measured cross sections) of the $^{169}$Tm($\alpha$,n)$^{172}$Lu reaction; recently measured entries are indicated by bold numbers.}
\begin{tabular}{cccccc}
\hline
\multicolumn{1}{c}{$E_{\alpha}$} &
\multicolumn{1}{c}{$E_\mathrm{c.m.}$} &
\multicolumn{1}{c}{$E_\mathrm{S}$} &
\multicolumn{1}{c}{$^{169}$Tm($\alpha$,n)$^{172}$Lu} &
\multicolumn{1}{c}{$^{169}$Tm($\alpha$,n)$^{172}$Lu} \\
\multicolumn{1}{c}{} &
\multicolumn{1}{c}{} &
\multicolumn{1}{c}{} & &
\multicolumn{1}{c}{recent data} \\
\multicolumn{1}{c}{(MeV)} &
\multicolumn{1}{c}{(MeV)} &
\multicolumn{1}{c}{(MeV)} &
\multicolumn{1}{c}{(10$^{28}$ MeV barn)} &
\multicolumn{1}{c}{ ($\mu$barn)} \\

\hline
11.5\footnotemark[1]  & $11.21 \pm 0.057$ & &     &\\
&&11.153&$77.210 \pm 8.127$     &\\
&&11.267&$51.778 \pm 5.450$ &\\

11.85\footnotemark[1] & $11.55 \pm 0.059$ &&     &\\
&&11.491&$62.822 \pm 6.746$ &\\
&&11.609&$42.309 \pm 4.543$ &\\

12.2\footnotemark[1]  & $11.90 \pm 0.061$ &&     &\\
&&11.839&$54.544 \pm 6.150$ &\\
&&11.961&$36.906 \pm 4.161$ &\\

12.5\footnotemark[1]  & $12.19 \pm 0.062$ &&     &\\
&&12.128&$42.697 \pm 3.799$ &\\
&&12.252&$29.116 \pm 2.591$ &\\

&&&&\\
\bf{12.6\footnotemark[2]}& $\bf{12.28 \pm 0.087}$   & &   & $\bf{9.82 \pm 0.92}$ \\
&&\bf{12.193}&$\bf{42.956 \pm 4.024}$&\\
&&\bf{12.367}&$\bf{25.252 \pm 2.366}$&\\

\bf{13.0\footnotemark[2]} & $\bf{12.68  \pm 0.066}$   & &  & $\bf{27.5 \pm 2.61}$\\
&&\bf{12.614}&$\bf{33.907 \pm 3.218}$&\\
&&\bf{12.746}&$\bf{23.096 \pm 2.192}$&\\

&&&&\\
13.5\footnotemark[3]\footnotemark[4]  & $13.16 \pm 0.066$ &&   &   \\
&&13.094&$30.076 \pm 1.984$&\\
&&13.226&$20.921 \pm 1.380$&\\

14.0\footnotemark[1]\footnotemark[4]  & $13.66 \pm 0.069$ & &   &  \\
&&13.591&$21.484 \pm 1.638$&\\
&&13.729&$15.010 \pm  1.144 $&\\

15.0\footnotemark[1]\footnotemark[4]  & 14.63 $\pm$ 0.075 &&    & \\
&&14.555&$13.789 \pm  1.087 $&\\
&&14.705&$ 9.704 \pm  0.765 $&\\

15.5\footnotemark[1]  & 15.12 $\pm$ 0.077 &&     &  \\
&&15.043&$10.894 \pm  0.892 $&\\
&&15.197&$ 7.730 \pm  0.633 $&\\

16.0\footnotemark[1]\footnotemark[4]  & 15.61 $\pm$ 0.079 &&    & \\
&&15.531&$ 6.922 \pm  0.546 $&\\
&&15.689&$ 4.949 \pm  0.390 $&\\

16.5\footnotemark[1]  & 16.10 $\pm$ 0.081 &&    & \\
&&16.019&$ 5.149 \pm  0.376 $&\\
&&16.181&$ 3.708 \pm  0.271 $&\\

17.0\footnotemark[1]  & 16.59 $\pm$ 0.084 &&    &  \\
&&16.506&$ 3.731 \pm  0.276 $&\\
&&16.674&$ 2.695 \pm  0.199 $&\\

17.5\footnotemark[1]  & 17.08 $\pm$ 0.086 &&  &   \\
&&16.994&$ 2.537 \pm  0.188 $&\\
&&17.166&$ 1.845 \pm  0.137 $&\\

\hline
\end{tabular}
\footnotetext{taken from \cite{kis_plb, kis_npa}}
\footnotetext{this work}
\footnotetext{remeasured in this work; average between the remeasured value and the one from \cite{kis_plb}, weighted by the statistical uncertainties; supersedes \cite{kis_plb}}
\footnotetext{Average values weighted by the statistical uncertainties are given when two irradiations were carried out at the same energy.}
\end{table}
\endgroup

\begingroup
\squeezetable
\begin{table}
\caption{\label{tab:results_ag} Extended experimental $S$ factors (and recently measured cross sections) of the $^{169}$Tm($\alpha$,$\gamma$)$^{173}$Lu reaction; recently measured entries are indicated by bold numbers.}
\begin{tabular}{cccccc}

\hline
\multicolumn{1}{c}{$E_{\alpha}$} &
\multicolumn{1}{c}{$E_\mathrm{c.m.}$} &
\multicolumn{1}{c}{$E_\mathrm{S}$} &
\multicolumn{1}{c}{$^{169}$Tm($\alpha$,$\gamma$)$^{173}$Lu} &
\multicolumn{1}{c}{$^{169}$Tm($\alpha$,$\gamma$)$^{173}$Lu} \\
\multicolumn{1}{c}{} &
\multicolumn{1}{c}{} &
\multicolumn{1}{c}{} &&
\multicolumn{1}{c}{recent data} \\
\multicolumn{1}{c}{(MeV)} &
\multicolumn{1}{c}{(MeV)} &
\multicolumn{1}{c}{(MeV)} &
\multicolumn{1}{c}{(10$^{25}$ MeV barn)} &
\multicolumn{1}{c}{ ($\mu$barn)} \\

\hline
\bf{12.6\footnotemark[2]}& $\bf{12.28 \pm 0.087}$   & &   & $\bf{0.60 \pm 0.15}$ \\
&&  \bf{12.193}&$\bf{2624.6 \pm  656.2} $&\\
&&  \bf{12.367}&$\bf{1542.9 \pm  385.7} $&\\

\bf{13.0\footnotemark[2]} & $\bf{12.68  \pm 0.066}$    & & & $\bf{0.87 \pm 0.13}$\\
&&  \bf{12.614}&$\bf{1072.7 \pm  160.3} $&\\
&&  \bf{12.746}&$ \bf{730.7 \pm  109.2} $&\\

&&&&\\
13.5\footnotemark[3]\footnotemark[4]\footnotemark[5]  & $13.16 \pm 0.066$ & &   & \\
&&  13.094&$ 401.850 \pm   40.812 $&\\
&&  13.226&$ 279.530 \pm   28.390 $&\\
&&&&\\
14.0\footnotemark[1]\footnotemark[4]  & $13.66 \pm 0.069$ & &   &  \\
&&  13.591&$ 213.200 \pm   18.109 $&\\
&&  13.729&$ 148.950 \pm   12.652 $&\\

15.0\footnotemark[1]\footnotemark[4]  & $14.63 \pm 0.075$ & &    & \\
&&  14.555&$  61.002 \pm    5.995 $&\\
&&  14.705&$  42.929 \pm    4.219 $&\\

15.5\footnotemark[1]  & $15.12 \pm 0.077$ & &    &  \\
&&  15.043&$  35.574 \pm    3.655 $&\\
&&  15.197&$  25.241 \pm    2.593 $&\\

16.0\footnotemark[1]\footnotemark[4]  & $15.61 \pm 0.079$ & &    & \\
&&  15.531&$  18.394 \pm    1.421 $&\\
&&  15.689&$  13.152 \pm    1.016 $&\\

16.5\footnotemark[1]  & $16.10 \pm 0.081$ &    & \\
&&  16.019&$   8.986 \pm    0.694 $&\\
&&  16.181&$   6.471 \pm    0.499 $&\\

17.0\footnotemark[1]  & $16.59 \pm 0.084$ & &    &  \\
&&  16.506&$   4.689 \pm    0.376 $&\\
&&  16.674&$   3.387 \pm    0.272 $&\\

&&&&\\
17.5\footnotemark[5]  & $17.08 \pm 0.086$ & & &    \\
&&  16.994&$   2.679 \pm    0.198 $&\\
&&  17.166&$   1.948 \pm    0.144 $&\\
\hline
\end{tabular}
\footnotetext{taken from \cite{kis_plb, kis_npa}}
\footnotetext{this work}
\footnotetext{remeasured in this work; average between the remeasured value and the one from \cite{kis_plb}, weighted by the statistical uncertainties; supersedes \cite{kis_plb}}
\footnotetext{Average values weighted by the statistical uncertainties are given when two irradiations were carried out at the same energy.}
\footnotetext{counted at LNGS; the average (weighted by the statistical uncertainty and the uncertainty of the detector efficiency) of the measured values is given}
\end{table}
\endgroup

The first results on the studied $^{169}$Tm+$\alpha$ reactions have been published already in abbreviated form \cite{kis_plb} and a full description of the experimental technique has been presented in \cite{kis_npa}. Recently, the shielding around the LEPS detector has been refined and thickened (details can be found in \cite{kis11_cgs}) this way the average laboratory background count rate is reduced to about 1.7 x 10$^{-3}$ 1/ (keV s) at the 20 - 80 keV energy region. Using this improved shielding it was possible to study the $^{169}$Tm($\alpha$,$\gamma$)$^{173}$Lu reaction at even lower energies, at $E_{\alpha}=13.0$ and 12.6 MeV. The experimental approach was similar to the one published in \cite{kis_plb, kis_npa}, here only additional information relevant for the recently measured cross sections is given.

The thicknesses of the thulium targets -- produced via vacuum evaporation onto 2 $\mu$m thick high purity Al backings -- were 331 and 377 $\mu$g/cm$^2$, corresponding to 1.18 x 10$^{18}$ and 1.34 x 10$^{18}$ atom/cm$^2$, respectively. The number of target atoms has been derived using weighing and the PIXE method. A single irradiation using a 13.0 MeV He$^{++}$ beam provided by the cyclotron accelerator of ATOMKI was carried out, the total number of the He$^{++}$ particles impinging on the targets was 8.76 x 10$^{17}$. Similarly to \cite{gyu06,yal09} an Al energy degrader foil was placed between the two thulium targets, the degrader foil thickness was determined by measuring the energy loss of alpha particles emitted by a $^{241}$Am radioactive source and by weighing. The energy losses in the thulium layer, in the backing and in the energy degrader foil were calculated using the SRIM code \cite{SRIM}.

After the end of the irradiation the activity of the two samples has been measured with the LEPS detector in far geometry to determine the $^{169}$Tm($\alpha$,n)$^{172}$Lu cross section by counting the yield of the emitted characteristic K$_{\alpha 1-2}$ X-rays. The length of these countings was about a day and were repeated once more about 3-4 days later. The X-ray counting was carried out again in close geometry about 14-17 weeks later to determine the number of the $^{173}$Lu isotopes produced in the  $^{169}$Tm($\alpha,\gamma$)$^{173}$Lu reaction. During the cooling period of more than 14 weeks the $^{172}$Lu activity of the targets decreased by a factor of about 16000, therefore the observed X-ray yield belongs solely to the decay of the $^{173}$Lu. The length of these countings was about 3 weeks.

The experimental data are shown in Tables \ref{tab:results_an} and \ref{tab:results_ag}. Consistent results were obtained for the cases of two irradiations at the same energy. At these energies the average cross section values weighted by the statistical uncertainty are given. Furthermore, to test the results based on X-ray counting, the activity of two samples were measured at the LNGS deep underground laboratory \cite{kis_npa}. For these irradiations, the average weighted by the uncertainty of the measured cross section values is given here. The quoted uncertainty in the $E_\mathrm{c.m.}$ values corresponds to the energy stability of the beam and to the uncertainty of the energy loss in the target. The uncertainty of the cross section is the quadratic sum of the following partial errors: efficiency of the HPGe and LEPS detectors (6 and 4 \%, respectively), number of target atoms (4\%), current measurement (3\%), uncertainty of decay parameters ($\leq$\ 5 \%) and counting statistics (0.5 - 13\%). The uncertainties given for the averaged values are the variances of the weighted means.

In astrophysical investigations it is common to quote the astrophysical $S$ factor. The cross section $\sigma(E)$ and the astrophysical $S$ factor $S(E)$ at c.m.\ energy $E_\mathrm{c.m.}$ are related by
\begin{equation}
 S(E_\mathrm{c.m.})=E_\mathrm{c.m.}\sigma(E_\mathrm{c.m.}) e^{2\pi\eta}\quad,
\end{equation}
with $\eta$ being the Sommerfeld parameter
\begin{equation}
\eta=\frac{Z_\mathrm{p}Z_\mathrm{T}e^2}{\hbar}\left( \frac{\mu}{2E_\mathrm{c.m.}} \right)^{1/2} \quad.
\end{equation}
The charge numbers $Z_\mathrm{p}$, $Z_\mathrm{T}$ of projectile and target, respectively, and their reduced mass $\mu$ enter the Sommerfeld parameter.
Since the energy enters the calculation of the $S$ factor also via the Sommerfeld parameter, the inclusion of the errors on cross sections and energies is not straightforward when computing the $S$ factor from measured cross sections. Because of this, a range of $S$ factors has to be given for each c.m.\ energy $E_\mathrm{S}$ in Tables \ref{tab:results_an} and \ref{tab:results_ag}, evaluated at the lower and upper limit of the energy range defined by the errors on $E_\mathrm{c.m.}$. The error bars on $\alpha$ energy and cross section translate into an error region for the $S$ factor which is of trapezoid shape, with its four corners given by the upper and lower limit of the $S$ factor at each energy $E_\mathrm{S}$. This error trapezoid is also shown in Figs.\ \ref{fig:ags} and \ref{fig:ans}.

\section{Discussion of implications for the astrophysical reaction rate}
\label{sec:dis}

\subsection{Relevant energy range and sensitivities}
\label{sec:energyrange}

In principle, reaction data can be used to test predictions of the cross sections and the resulting reaction rates for astrophysics in two ways. A direct comparison to data is useful if they reach the astrophysical energy range or, at least, are taken at energies where the cross sections show a similar sensitivity to the reaction widths as at astrophysically relevant energies. If such data are not available, the combined data of different reaction channels may be used to extract information on the quantities determining the astrophysical reaction rates.

\begin{figure}
\includegraphics[angle=-90,width=\columnwidth]{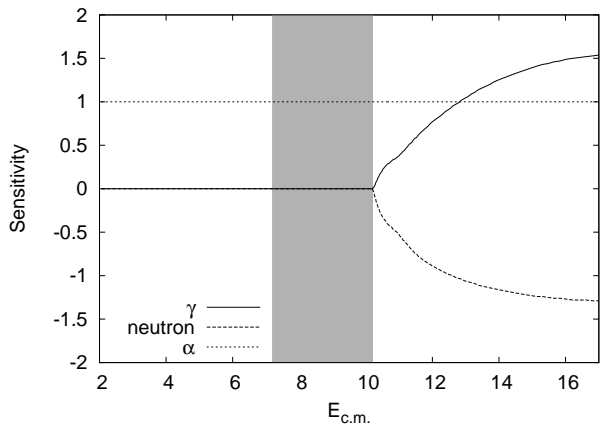}
\caption{\label{fig:agsensi}Sensitivity of the reaction $^{169}$Tm($\alpha$,$\gamma$)$^{173}$Lu to a variation of $\gamma$-, neutron-, and $\alpha$-widths (variation of proton widths is not shown as the cross sections are insensitive to it) as function of c.m.\ energy in MeV; the astrophysically relevant energy range $7.2-10.2$ MeV at $T=2.5$ GK is shown by the shaded region.}
\end{figure}

The astrophysical energy window for $^{169}$Tm($\alpha$,$\gamma$) is $6.3-9.2$ MeV at 2 GK and $7.8-10.7$ MeV at 3 GK \cite{energywindows}. The present data come close to the energy window but do not reach it. The reaction rate is only sensitive to the $\alpha$ width because it is the smallest width in the relevant energy region due to the Coulomb barrier \cite{sensi}. The sensitivity of the reaction cross section, shown in Fig.\ \ref{fig:agsensi}, depends complicatedly on energy. For a discussion of the sensitivity definition, see \cite{raureview,sensi}. The sensitivity $s$ is defined in such a way that $|s|=1$ implies a change in the cross section by the same factor as the width was changed. When $s<1$ the cross section values change opposite to the variation, i.e., they decrease when the corresponding width is increased.

Below the neutron threshold, the ($\alpha$,$\gamma$) reaction is only sensitive to a change in the $\alpha$ width. Therefore it is sensitive to all uncertainties involving the optical $\alpha$+nucleus potential. Above the threshold, it is still sensitive to the $\alpha$ width but also uncertainties in the neutron- and $\gamma$-widths become increasingly important with increasing energy. At $E_\mathrm{c.m.}>12.8$ MeV, the sensitivity to these quantities exceeds the one for the $\alpha$ width. Sensitivities larger than one appear because $\alpha$ particles may be re-emitted in the $\gamma$ cascade when the compound nucleus de-excites \cite{sensi}. Such emissions are only included approximately in the version of the reaction code used and therefore larger uncertainties in the predictions may be expected at those energies.

\begin{figure}
\includegraphics[angle=-90,width=\columnwidth]{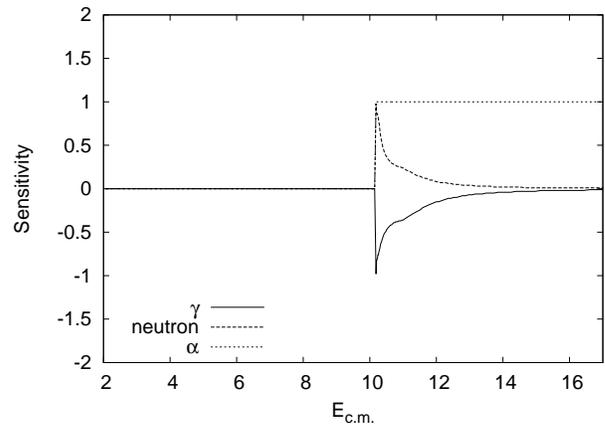}
\caption{\label{fig:ansensi}Sensitivity of the reaction $^{169}$Tm($\alpha$,n)$^{172}$Lu to a variation of $\gamma$-, neutron-, and $\alpha$-widths (variation of proton widths is not shown as the cross sections are insensitive to it) as function of c.m.\ energy in MeV; the shown energy range was deliberately chosen to be the same as in Fig.\ \ref{fig:agsensi} to facilitate a comparison.}
\end{figure}

The energy range $12.19\leq E_\mathrm{c.m.}\leq 17.17$ MeV of the measured ($\alpha$,$\gamma$) cross sections covers a region where three different widths are important. The situation is further complicated by the fact that in this reaction with highly negative $Q$ values in all channels, additional reaction channels appear, such as the cascade emission of $\alpha$ particles. In consequence, it is hard to disentangle the different contributions to test the astrophysically important quantity, the predicted $\alpha$ width. Cross comparison with the $^{169}$Tm($\alpha$,n)$^{172}$Lu data helps in this task. Figure \ref{fig:ansensi} shows that the ($\alpha$,n) reaction is mainly sensitive to the $\alpha$ width at the higher energies. Only close to the threshold, a stronger sensitivity to changes in the neutron- and $\gamma$-widths appears.


\subsection{Comparison of theory to data}
\label{sec:theodata}

\begin{figure}
\includegraphics[angle=-90,width=\columnwidth]{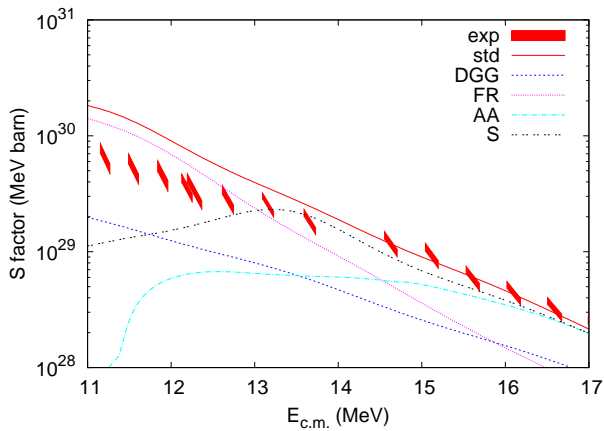}
\caption{\label{fig:anpots}(color online) Experimental astrophysical $S$ factors (exp) of $^{169}$Tm($\alpha$,n)$^{172}$Lu as function of c.m.\ energy are compared to values calculated with the global $\alpha$+nucleus potentials by \cite{mcf} (std), \cite{demet} (DGG), \cite{frodip,raufro} (FR), \cite{avri} (AA), and \cite{sauer11} (S) .}
\end{figure}

\begin{figure}
\includegraphics[angle=-90,width=\columnwidth]{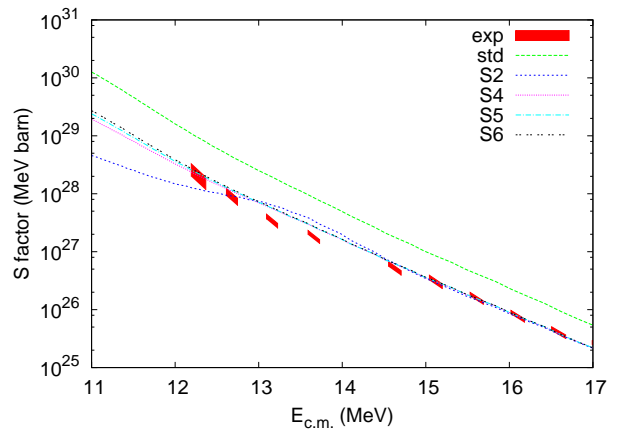}
\caption{\label{fig:ags}(color online) Experimental astrophysical $S$ factors (exp) of $^{169}$Tm($\alpha$,$\gamma$)$^{173}$Lu as function of c.m.\ energy are compared to values calculated with different inputs; see text for further details.}
\end{figure}

\begin{figure}
\includegraphics[angle=-90,width=\columnwidth]{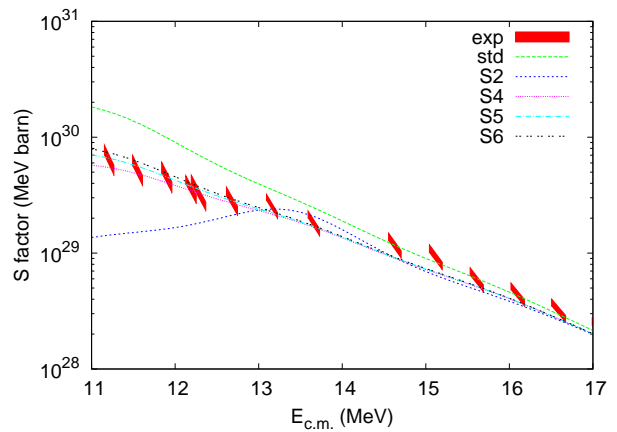}
\caption{\label{fig:ans}(color online) Same as Fig.\ \ref{fig:ags} but for $^{169}$Tm($\alpha$,n)$^{172}$Lu}
\end{figure}

Armed with the insights from the previous section, we can interpret the comparison between experiment and theory shown in Figs.\ \ref{fig:anpots} -- \ref{fig:ans} for the ($\alpha$,n) and ($\alpha$,$\gamma$) reactions. The experimental $S$ factors in these figures are compared to theoretical values obtained with various settings of the statistical model code SMARAGD, version 0.8.4s \cite{raureview,SMARAGD}. It is important to mention that SMARAGD uses an improved routine to solve the Schr\"odinger equation, e.g., compared to the code used in \cite{adndt,adndt1}, leading to different results especially at energies close to the Coulomb barrier (see \cite{raureview}). The calculated SMARAGD standard value (labeled 'std') uses the optical potential by \cite{mcf}.
It perfectly reproduces the ($\alpha$,n) data above 14 MeV, as seen in Fig.~\ref{fig:anpots}. For completeness, the $S$ factors obtained with more recent global $\alpha$+nucleus optical potentials from \cite{demet,frodip,raufro,avri,sauer11} are also shown. They exhibit a very different energy dependence and cannot even describe the measured data above 14 MeV, with the exception of the potential by \cite{sauer11} which uses an energy-dependent parameterization approaching the parameters of \cite{mcf} at high energies (see Sec.~\ref{sec:extension}). Therefore, the following discussion of ($\alpha$,$\gamma$) predictions and a possible extension to astrophysical energies focusses on the optical potential of \cite{mcf} and variations of the potential from \cite{sauer11}.

Regarding the ($\alpha$,$\gamma$) reaction, the standard theory values exceed the experimental $S$ factors by factors of 2.5 and four at the upper and lower end, respectively, of the measured energy range (Fig.~\ref{fig:ags}). Apparently, not only is the absolute value not reproduced but also the energy dependence is different. Comparing the prediction to the ($\alpha$,n) data in Figs.\ \ref{fig:anpots} and \ref{fig:ans}, however, shows that a mispredicted $\alpha$ width cannot be the culprit for the deviation at energies above 14 MeV. As mentioned above, theory agrees with the experimental $S$ factor values at these energies. Below 14 MeV, on the other hand, a different energy dependence is seen in the ($\alpha$,n) data when compared to theory. At first glance, this may be attributed to a problem in the prediction of the neutron- or $\gamma$-width because the $S$ factor is increasingly sensitive to these widths towards lower energy, as depicted in Fig.~\ref{fig:ansensi}. In fact, either an increase in the $\gamma$ width or a decrease in the neutron width by a factor of 5 allows to reproduce the ($\alpha$,n) data across the full energy range.

It has to be realized, however, that the sensitivities to neutron and $\gamma$ widths are opposite in the ($\alpha$,n) and ($\alpha$,$\gamma$) reactions. In consequence, any attempt to remove the discrepancy of theory with the ($\alpha$,n) data by modifying the neutron and/or $\gamma$ channel will result in shifting the calculated ($\alpha$,$\gamma$) $S$ factors to higher values, further away from the experimental ones. Using the above modification factor of 5, the new theory prediction for ($\alpha$,$\gamma$) would be a factor of 20 above the experimental values although the energy dependence is then reproduced well.
Therefore, we have to proceed differently. The ($\alpha$,n) data prove that the $\alpha$ width is described well above 14 MeV. The deviations below 14 MeV, nevertheless, may still be partly due to an incorrect energy dependence of the $\alpha$ width. From the ($\alpha$,$\gamma$) data, a factor of 2.5 between theory and experiment is extracted from the values above 14 MeV. This factor must come from incorrectly predicted $\gamma$- or neutron-widths. It is inconclusive with the given data which of the widths is incorrect and therefore we can compensate this factor by either increasing the neutron width or decreasing the $\gamma$ width. This will be of no consequence for the astrophysical rate which only depends on the $\alpha$ width. After having modified the neutron or $\gamma$ width for a given $\alpha$ width so that it reproduces the ($\alpha$,$\gamma$) $S$-factors above 14 MeV, remaining deviations from both ($\alpha$,$\gamma$) and ($\alpha$,n) data must be due to the $\alpha$ width. The investigations discussed in the following and regarding the $\alpha$+nucleus optical potential have been based on a $\gamma$ width renormalized by a factor 0.5. All theoretical $S$ factor curves shown in Figs.~\ref{fig:ags}, \ref{fig:ans}, and \ref{fig:rateratio} -- except for the 'std' values -- include the modified $\gamma$ width.


\subsection{Extension to astrophysical energies}
\label{sec:extension}

An energy-dependent optical $\alpha$+nucleus potential, which transforms into the successful potential of \cite{mcf} at projectile energies well above the Coulomb barrier but has a shallower imaginary part at low energy, was suggested in \cite{sauer11}. The potential uses the same real part and the same geometry as the potential of \cite{mcf} but the strength of the volume imaginary part is parameterized with respect to the Coulomb barrier $E_\mathrm{C}$
\begin{equation}
\label{eq:w}
W=\frac{25}{1+e^{\left(0.9E_\mathrm{C}-E_\mathrm{c.m.} \right)/a_\mathrm{C}}} \quad \mathrm{MeV},
\end{equation}
with $a_\mathrm{C}=2$ MeV. At high energy $E_\mathrm{c.m.}$ in the $\alpha$ channel, the depth $W$ will assume the standard value from \cite{mcf}.

The $S$ factor curves obtained with this potential are labeled 'S2' in Figs.~\ref{fig:ags}, \ref{fig:ans}. The energy dependence of the potential is too strong when applied to this reaction, as can be seen in Fig.~\ref{fig:ans}, although it fared well in describing the $^{141}$Pr($\alpha$,n)$^{144}$Pm data of \cite{sauer11}. The only free parameter is $a_\mathrm{C}$, determining the strength of the energy dependence. The curves labeled 'S4', 'S5', and 'S6' use $a_\mathrm{C}$ values of 4, 5, and 6 MeV, respectively. All can describe the present $^{169}$Tm($\alpha$,n) data well and simultaneously also the capture data, the latter being shown in Fig.~\ref{fig:ags}.

\begin{figure}
\includegraphics[angle=-90,width=\columnwidth]{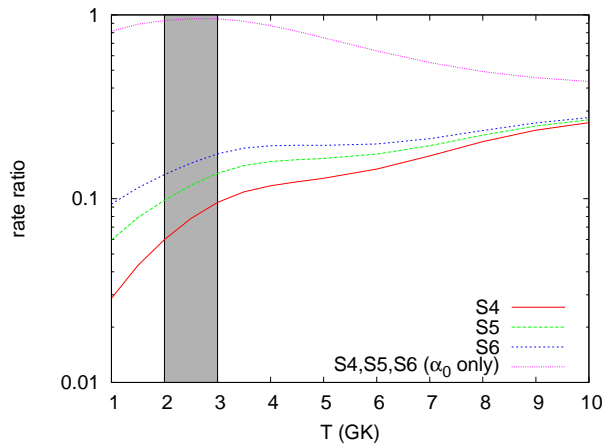}
\caption{\label{fig:rateratio}(color online) Ratios $r^*_{\mathrm{S}x}/r^*_\mathrm{std}$ of the astrophysical $^{169}$Tm($\alpha$,$\gamma$)$^{173}$Lu reaction rates $r^*_{\mathrm{S}x}$ calculated with different potentials S$x$ ($x=4,5,6$) to the rate $r^*_\mathrm{std}$ calculated with the standard potential by \cite{mcf} as function of plasma temperature $T$; the temperature range relevant in the $\gamma$ process is marked by the shaded region.}
\end{figure}

It should be noted that using the above approach, the optical potential \textit{effectively} includes everything which may alter the $\alpha$ width at low energy. The $\alpha$ width can be affected by the included low-lying levels and the optical potential \cite{raureview,sensi}.
The level scheme of $^{169}$Tm at low excitation energy is well established and can be assumed to be sufficiently known. This leaves the possibility of an incorrect energy dependence of the optical potential. Another possibility, however, would be an alteration of the compound formation cross section by direct processes. The formation cross section is determined by the $\alpha_0^\mathrm{cap}$ width (i.e.\ the width for resonant $\alpha$ capture on the ground state (g.s.) of the target nucleus) and affects all compound reaction channels. An optical potential is usually derived from fits to elastic scattering data. Its imaginary part describes the flux into non-elastic channels but without distinguishing the nature of these channels. If such direct processes were implicitly included in the standard optical potential, the formation cross section would be overestimated because it is implicitly assumed in the reaction calculation that all $\alpha$ flux missing from the elastic channel will contribute to the formation of a compound nucleus. Therefore, the modification of the optical $\alpha$+nucleus potential suggested above has to be viewed as an \textit{effective} correction, ignoring the cause for the correction. If some sort of direct process was acting at low energy, explicit inclusion of such a process in the reaction calculation would remove the need to alter the optical potential.

Assuming that Eq.~\eqref{eq:w} is the correct function for describing the energy dependence, the limits on the astrophysical reaction rates for $4\leq a_\mathrm{C} \leq 6$ MeV can be calculated at plasma temperatures relevant for the $\gamma$ process.
The astrophysical reaction rates resulting from the use of the three potentials are compared to the rate obtained with the standard potential in Fig.~\ref{fig:rateratio}. The new rates are at least a factor of 5 below our standard rate obtained with the potential of \cite{mcf}. At $T=2$ GK, the ratios between newly calculated rates and our standard prediction range from 0.06 to 0.136 when increasing $a_\mathrm{C}$ from 4 to 6 MeV.

Table \ref{tab:newrate} gives the stellar reactivities $N_A \left< \sigma v \right>^*$ for the reactions $^{169}$Tm($\alpha$,$\gamma$)$^{173}$Lu and $^{173}$Lu,($\gamma$,$\alpha$)$^{169}$Tm computed with $a_\mathrm{C}=4, 5, 6$ MeV at different plasma temperatures. Table \ref{tab:ratefit} presents the coefficients for the usual REACLIB parameterization \cite{adndt}. Due to the negative reaction $Q$ value, the parameter fits had to be treated specially. They were obtained by converting the calculated stellar ($\alpha$,$\gamma$) reactivity to the one for stellar ($\gamma$,$\alpha$), using the reciprocity relations between forward and reverse \textit{stellar} reactivities \cite{raureview,fow74}. This was then fitted and the parameters were converted back to the one for the ($\alpha$,$\gamma$) reaction as shown in \cite{adndt}. This ensures better numerical accuracy than fitting a reaction with negative $Q$ value. %The reactivities obtained with $a_\mathrm{C}=4$ MeV and $a_\mathrm{C}=6$ MeV can be viewed as lower and upper limits, respectively.

\begin{table*}
\caption{\label{tab:newrate}Stellar reactivity $N_A \left< \sigma v \right>^*$ for $^{169}$Tm($\alpha$,$\gamma$)$^{173}$Lu as function of plasma temperature $T$ obtained with three different values for $a_\mathrm{C}$ -- 4, 5, and 6 MeV -- labeled by S4, S5, and S6, respectively.}
\begin{ruledtabular}
\begin{tabular}{rllll}
\multicolumn{1}{c}{$T$} & \multicolumn{1}{c}{S4} & \multicolumn{1}{c}{S5} & \multicolumn{1}{c}{S6} & \multicolumn{1}{c}{S6 ($\alpha_0$ only)} \\
\multicolumn{1}{c}{(GK)} & \multicolumn{1}{c}{(cm$^3$s$^{-1}$mole$^{-1}$)} & \multicolumn{1}{c}{(cm$^3$s$^{-1}$mole$^{-1}$)} & \multicolumn{1}{c}{(cm$^3$s$^{-1}$mole$^{-1}$)} & \multicolumn{1}{c}{(cm$^3$s$^{-1}$mole$^{-1}$)} \\
\hline
0.50  & $ 1.009\times 10^{-52 }$ & $ 2.600\times 10^{-52 }$ & $ 4.557\times 10^{-52}$& $ 4.600\times 10^{-51 }$ \\
0.60  & $ 2.600\times 10^{-47 }$ & $ 6.301\times 10^{-47 }$ & $ 1.086\times 10^{-46}$& $ 1.059\times 10^{-45 }$ \\
0.70  & $ 4.942\times 10^{-43 }$ & $ 1.147\times 10^{-42 }$ & $ 1.923\times 10^{-42}$& $ 1.817\times 10^{-41 }$ \\
0.80  & $ 1.550\times 10^{-39 }$ & $ 3.454\times 10^{-39 }$ & $ 5.644\times 10^{-39}$& $ 5.183\times 10^{-38 }$ \\
0.90  & $ 1.308\times 10^{-36 }$ & $ 2.807\times 10^{-36 }$ & $ 4.488\times 10^{-36}$& $ 4.014\times 10^{-35 }$ \\
1.00  & $ 4.089\times 10^{-34 }$ & $ 8.502\times 10^{-34 }$ & $ 1.332\times 10^{-33}$& $ 1.162\times 10^{-32 }$ \\
1.50  & $ 1.594\times 10^{-25 }$ & $ 2.902\times 10^{-25 }$ & $ 4.214\times 10^{-25}$& $ 3.270\times 10^{-24 }$ \\
2.00  & $ 2.522\times 10^{-20 }$ & $ 4.129\times 10^{-20 }$ & $ 5.693\times 10^{-20}$& $ 3.920\times 10^{-19 }$ \\
2.50  & $ 9.190\times 10^{-17 }$ & $ 1.392\times 10^{-16 }$ & $ 1.838\times 10^{-16}$& $ 1.122\times 10^{-15 }$ \\
3.00  & $ 3.733\times 10^{-14 }$ & $ 5.371\times 10^{-14 }$ & $ 6.857\times 10^{-14}$& $ 3.721\times 10^{-13 }$ \\
3.50  & $ 3.533\times 10^{-12 }$ & $ 4.911\times 10^{-12 }$ & $ 6.111\times 10^{-12}$& $ 2.995\times 10^{-11 }$ \\
4.00  & $ 1.157\times 10^{-10 }$ & $ 1.565\times 10^{-10 }$ & $ 1.909\times 10^{-10}$& $ 8.594\times 10^{-10 }$ \\
4.50  & $ 1.707\times 10^{-9 }$ & $ 2.250\times 10^{-9 }$ & $ 2.697\times 10^{-9}$   & $ 1.124\times 10^{-8 }$  \\
5.00  & $ 1.385\times 10^{-8 }$ & $ 1.777\times 10^{-8 }$ & $ 2.093\times 10^{-8}$   & $ 8.060\times 10^{-8 }$  \\
6.00  & $ 2.721\times 10^{-7 }$ & $ 3.283\times 10^{-7 }$ & $ 3.725\times 10^{-7}$   & $ 1.197\times 10^{-6 }$  \\
7.00  & $ 1.998\times 10^{-6 }$ & $ 2.271\times 10^{-6 }$ & $ 2.481\times 10^{-6}$   & $ 6.434\times 10^{-6 }$  \\
8.00  & $ 8.180\times 10^{-6 }$ & $ 8.874\times 10^{-6 }$ & $ 9.406\times 10^{-6}$   & $ 1.967\times 10^{-5 }$  \\
9.00  & $ 2.184\times 10^{-5 }$ & $ 2.302\times 10^{-5 }$ & $ 2.393\times 10^{-5}$   & $ 4.217\times 10^{-5 }$  \\
10.00  & $ 4.117\times 10^{-5 }$ & $ 4.272\times 10^{-5 }$ & $ 4.391\times 10^{-5}$  & $ 6.901\times 10^{-5 }$
\end{tabular}
\end{ruledtabular}
\end{table*}

\begingroup
\squeezetable
\begin{table*}
\caption{\label{tab:ratefit}REACLIB parameters obtained from fitting the reactivities shown in Table \ref{tab:newrate}}
\begin{ruledtabular}
\begin{tabular}{ccccccccc}
 & \multicolumn{2}{c}{S4} & \multicolumn{2}{c}{S5} & \multicolumn{2}{c}{S6} & \multicolumn{2}{c}{S6 ($\alpha_0$ only)} \\
          & \multicolumn{1}{c}{($\alpha$,$\gamma$)} & \multicolumn{1}{c}{($\gamma$,$\alpha$)} & \multicolumn{1}{c}{($\alpha$,$\gamma$)} & \multicolumn{1}{c}{($\gamma$,$\alpha$)} & \multicolumn{1}{c}{($\alpha$,$\gamma$)} & \multicolumn{1}{c}{($\gamma$,$\alpha$)} & \multicolumn{1}{c}{($\alpha$,$\gamma$)} & \multicolumn{1}{c}{($\gamma$,$\alpha$)} \\
\hline
%$a_0$ & -2.380584\times 10^{2} & 2. & 3. & 4. & 5. & 6.
$a_0$  & $-2.380584\times 10^{2}$ & $-2.143877\times 10^{2}$ & $-2.525444\times 10^{2}$ & $-2.288737\times 10^{2}$ & $-2.134617\times 10^{2}$ & $-1.897910\times 10^{2 }$ & $-1.521031\times 10^{2}$ & $-1.284324\times 10^{2}$ \\
$a_1$  & $-2.285393\times 10^{1}$ & 0.000000 & $-2.285393\times 10^{1}$ & 0.000000 & $-2.285393\times 10^{1}$ &  0.000000 & $-2.285393\times 10^{1}$ & 0.000000 \\
$a_2$  & \multicolumn{2}{c}{$-8.573430\times 10^{1}$} &  \multicolumn{2}{c}{$-1.132236\times 10^{2}$} & \multicolumn{2}{c}{$-6.521460\times 10^{1}$} & \multicolumn{2}{c}{0.000000}  \\
$a_3$  &  \multicolumn{2}{c}{$2.974330\times 10^{2}$} & \multicolumn{2}{c}{$3.424164\times 10^{2}$} & \multicolumn{2}{c}{$2.499602\times 10^{2}$} & \multicolumn{2}{c}{$1.174915\times 10^{2}$}  \\
$a_4$  & \multicolumn{2}{c}{$-2.842263\times 10^{1}$} &  \multicolumn{2}{c}{$-3.081515\times 10^{1}$} & \multicolumn{2}{c}{$-2.456620\times 10^{1}$} & \multicolumn{2}{c}{$-1.591620\times 10^{1}$} \\
$a_5$  & \multicolumn{2}{c}{$1.740137$} & \multicolumn{2}{c}{$1.864776$} & \multicolumn{2}{c}{$1.437111$} & \multicolumn{2}{c}{$8.700436\times 10^{-1}$} \\
$a_6$  & $-7.211434\times 10^{1}$ & $-7.061434\times 10^{1}$ & $-9.447688\times 10^{1}$ & $-9.297688\times 10^{1}$ & $-5.338706\times 10^{1}$ &  $-5.188706\times 10^{1}$ & $4.644152$ & $6.144152$
\end{tabular}
\end{ruledtabular}
\end{table*}
\endgroup

\begin{figure}
\includegraphics[angle=-90,width=\columnwidth]{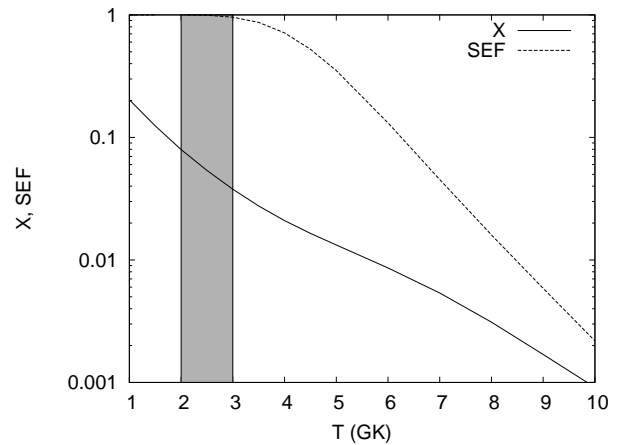}
\caption{\label{fig:xsef}Comparison of g.s.\ contribution $X$ (X) (taken from \cite{sensi}) and stellar enhancement factor $f_\mathrm{SEF}$ (SEF) of $^{169}$Tm($\alpha$,$\gamma$)$^{173}$Lu as function of plasma temperature. The $\gamma$-process temperature range is marked by the shaded area.}
\end{figure}

As discussed above, it has to be cautioned, however, that the presented reactivities still carry large uncertainties as the energy-dependence of the $\alpha$ width at energies close to the Coulomb barrier is still not well understood. Moreover, strictly speaking the experiment only provides an indication of how the $\alpha_0^\mathrm{cap}$ width is modified. Stellar reactivities include transitions from and to excited states of the target nucleus, too. Figure \ref{fig:rateratio} also shows the ratio of an ''$\alpha_0$ only'' rate to our comparison standard. The ''$\alpha_0$ only'' rate was obtained by using the potential modification of Eq.\ \eqref{eq:w} only for $\alpha$ transitions from and to the g.s.\ of $^{169}$Tm while using the standard potential for all others. Incidentally, this yields an almost unchanged rate at $\gamma$-process temperatures. (The results for different values of $a_\mathrm{C}$ are indistinguishable in the plot.) The shown behavior can be understood by realizing that the temperature dependence of the altered g.s.\ rate is folded with the contribution $X$ of the g.s.\ transitions to the total stellar rate. Figure \ref{fig:xsef} shows values for the g.s.\ contribution $X$ as function of temperature, taken from \cite{sensi}. It can be seen that $X$ is small at $\gamma$-process temperatures. Often, the stellar enhancement factor (SEF)
\begin{equation}
f_\mathrm{SEF}=\frac{\left< \sigma v \right>^*}{\left< \sigma v \right>^\mathrm{g.s.}}
\end{equation}
is quoted as a comparison between stellar reactivity and the reactivity computed from g.s.\ transitions alone. The SEF is unsuited, however, to judge the contribution of the laboratory cross section to the stellar rate, as discussed in \cite{xfactor}. The reaction $^{169}$Tm($\alpha$,$\gamma$)$^{173}$Lu is a good example for that. As found in Fig.~\ref{fig:xsef}, $f_\mathrm{SEF}\approx 1$ in the relevant temperature range whereas $X$ is much lower.

The stellar reactivities and the REACLIB parameters for the ''$\alpha_0$ only'' case and $a_\mathrm{C}=6$ MeV are also given in Tables \ref{tab:newrate} and \ref{tab:ratefit}, respectively. Without a further understanding of the nature of the $\alpha$ width modification, the upper limit of the stellar reactivity is given by the values of model S6 ($\alpha_0$ only) in the last column of Table \ref{tab:newrate}, while the lower limit is provided by the reactivities for S4 ($a_\mathrm{C}=4$ MeV).

%
% \begin{table}
% \caption{\label{tab:a0pars}REACLIB parameters obtained from fitting the reactivities of S6 ($\alpha_0$ only) shown in Table \ref{tab:newrate}}
% \begin{ruledtabular}
% \begin{tabular}{ccc}
% Parameter & \multicolumn{1}{c}{($\alpha$,$\gamma$)} & \multicolumn{1}{c}{($\gamma$,$\alpha$)} \\
% \hline
% %$a_0$ & -2.380584\times 10^{2} & 2. & 3. & 4. & 5. & 6.
% $a_0$  & $-1.521031\times 10^{2}$ & $-1.284324\times 10^{2}$ \\
% $a_1$  & $-2.285393\times 10^{1}$ & 0.000000  \\
% $a_2$  & \multicolumn{2}{c}{0.000000}  \\
% $a_3$  & \multicolumn{2}{c}{$1.174915\times 10^{2}$}   \\
% $a_4$  & \multicolumn{2}{c}{$-1.591620\times 10^{1}$}  \\
% $a_5$  & \multicolumn{2}{c}{$8.700436\times 10^{-1}$}  \\
% $a_6$  & $4.644152$ & $6.144152$
% \end{tabular}
% \end{ruledtabular}
% \end{table}
%

\section{Summary and conclusion}
\label{sec:con}

We have determined the reaction cross sections and astrophysical $S$ factors of $^{169}$Tm($\alpha$,n)$^{172}$Lu and $^{169}$Tm($\alpha$,$\gamma$)$^{173}$Lu at low energies, using our newly developed method of combining activation and X-ray counting. An improved shielding around the LEPS detector enabled us to measure the ($\alpha$,$\gamma$) reaction down to 12.6 MeV $\alpha$ energy, lower than before.

The impact of the new results on the determination of the astrophysical reaction rates for $\alpha$ capture on $^{169}$Tm were discussed, using the full range of data. The combination of ($\alpha$,n) and ($\alpha$,$\gamma$) data was essential to disentangle errors in the predicted $\alpha$ width from those in other widths appearing in the reaction. It was found that the energy dependence of the $\alpha$ width below 14 MeV is different from what is expected by standard predictions. A modified, energy-dependent, local optical $\alpha$+nucleus potential was presented, able to describe both ($\alpha$,n) and ($\alpha$,$\gamma$) $S$ factors well across the measured energy range.

Using the local potential, stellar ($\alpha$,$\gamma$) and ($\gamma$,$\alpha$) reactivities were calculated. They were found to be considerably lower at astrophysical $\gamma$-process temperatures than predictions using a standard optical potential. More speicifically, they were factors of $7.4-16.7$ below the rate calculated with the potential by \cite{mcf}. Ambiguities in the extrapolation to low energies, however, require an uncertainty of a factor of $2-3$ in the predicted rate, even when assuming that the shape of the energy dependence is understood. Further ($\alpha$,$\gamma$) measurements below the neutron threshold would be able to reduce this uncertainty.

A further problem remains with identifying the nature of the potential modification. This leads to an even larger uncertainty in the calculated rate. Again, more low-energy data and an extended database including a wider range of nuclei would help to shed light on this issue. In conclusion, the recommended reactivity at a $\gamma$-process temperature of 2 GK is $0.06-0.95$ times the SMARAGD reactivity using the potential by \cite{mcf}, which translates to $2.3-37.0$ times the widely used standard values of \cite{adndt1}, thereby leading to an \textit{enhancement} in the ($\gamma$,$\alpha$) rate with respect to the values given in \cite{adndt,adndt1}.

The case studied here is a good example for the restrictions and possible pitfalls that can be encountered when deriving astrophysical reaction rates from experimental data. It not only shows the importance of further measurements of reaction cross sections involving low-energy $\alpha$ particles to allow global studies of suitable optical potentials but also the significance of properly accounting for both, changing width sensitivities as well as thermally excited states, when interpreting the impact on an astrophysical reaction rate.

\section*{Acknowledgments}
This work was supported by the EUROGENESIS research
program, the European Research Council grant agreement no. 203175, the Economic Competitiveness Operative Programme GVOP-3.2.1.-2004-04-0402/3.0., OTKA (NN83261, K101328, PD104664), the ENSAR/THEXO European FP7 programme, and the T\'AMOP-4.2.2/B-10/1-2010-0024 project. This project is co-financed by the European Union and the European Social fund. CF~is supported by the DOE Topical Collaboration "Neutrinos and Nucleosynthesis in Hot and Dense Matter" under contract DE-FG02-10ER41677.

\end{document}